\documentclass[twocolumn, letterpaper, aps, prapplied, floatfix, 10pt]{revtex4-2}

\pdfoutput=1

\usepackage{amsmath}
\usepackage{amsthm}
\usepackage{amssymb}
\usepackage{booktabs}
\usepackage{graphicx}
\usepackage[colorlinks=true, linkcolor=blue, citecolor=blue, urlcolor=blue, pdfauthor={Graham J. Norris}, pdftitle={Improved Parameter Targeting in {3D}-Integrated Superconducting Circuits through a Polymer Spacer Process}]{hyperref}
\usepackage{mathtools}
\usepackage[version=4]{mhchem}
\usepackage{physics}
\usepackage{siunitx}
\usepackage{xcolor}

\DeclareMathOperator{\K}{K}

\DeclareSIUnit\sccm{sccm}
\DeclareSIUnit\mK{mK}

\begin{document}
	\title{Improved Parameter Targeting in {3D}-Integrated Superconducting Circuits\\through a Polymer Spacer Process}

	\author{Graham J. Norris}
	\email{graham.norris@phys.ethz.ch}
	\affiliation{Department of Physics, ETH Z{\"u}rich, CH-8093 Z{\"u}rich, Switzerland}

	\author{Laurent Michaud}
	\affiliation{Department of Physics, ETH Z{\"u}rich, CH-8093 Z{\"u}rich, Switzerland}

	\author{David Pahl}
	\affiliation{Department of Physics, ETH Z{\"u}rich, CH-8093 Z{\"u}rich, Switzerland}

	\author{Michael Kerschbaum}
	\affiliation{Department of Physics, ETH Z{\"u}rich, CH-8093 Z{\"u}rich, Switzerland}

	\author{Christopher Eichler}
	\affiliation{Department of Physics, ETH Z{\"u}rich, CH-8093 Z{\"u}rich, Switzerland}

	\author{Jean-Claude Besse}
	\affiliation{Department of Physics, ETH Z{\"u}rich, CH-8093 Z{\"u}rich, Switzerland}

	\author{Andreas Wallraff}
	\email{andreas.wallraff@phys.ethz.ch}
	\affiliation{Department of Physics, ETH Z{\"u}rich, CH-8093 Z{\"u}rich, Switzerland}

	\date{July 5, 2023}

	\begin{abstract}
        Three-dimensional device integration facilitates the construction of superconducting quantum information processors with more than several tens of qubits by distributing elements such as control wires, qubits, and resonators between multiple layers.
		The frequencies of resonators and qubits in flip-chip-bonded multi-chip modules depend on the details of their electromagnetic environment defined by the conductors and dielectrics in their vicinity.
		Accurate frequency targeting therefore requires precise control of the separation between chips and minimization of their relative tilt.
		Here, we describe a method to control the inter-chip separation by using polymer spacers.
		Compared to an identical process without spacers, we reduce the measured planarity error by a factor of \num{3.5}, to a mean tilt of \SI{76(35)}{\micro \radian}, and the deviation from the target inter-chip separation by a factor of ten, to a mean of \SI{0.4(8)}{\um}.
		We apply this process to coplanar waveguide resonator samples and observe chip-to-chip resonator frequency variations below \SI{50}{\MHz} ($\approx \SI{1}{\percent}$).
		We measure internal quality factors of \num{5e5} at the single-photon level, suggesting that the added spacers are compatible with low-loss device fabrication.
	\end{abstract}

	\maketitle

	\section{Introduction}

	\begin{figure*}
		\centering
		\includegraphics{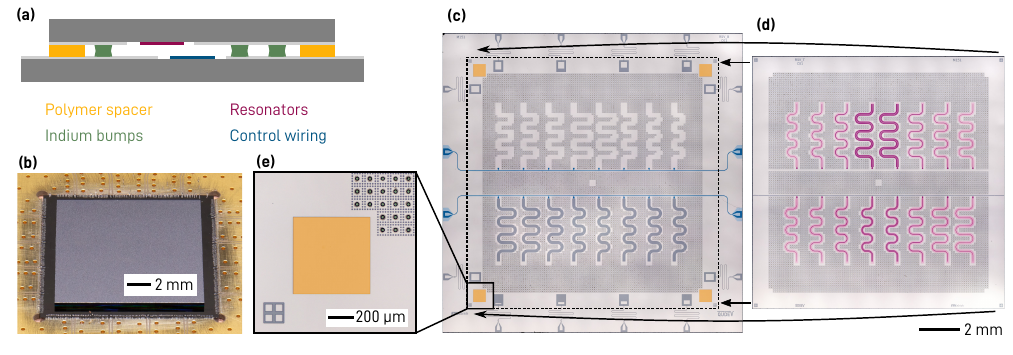}
		\caption{
			Overview of the presented 3D-integration scheme.
			(a) Schematic of our flip-chip bonding architecture including polymer spacers (not to scale).
			(b) Photograph of a flip-chip-bonded module wire-bonded to a PCB.
			(c) Colorized optical micrograph of the bottom wiring chip and (d) top resonator chip featuring feedlines (blue), resonators (red) and SU-8 spacers (yellow).
			(e) Detail of an SU-8 spacer also showing indium bumps (green).
		}
		\label{fig:main:1}
	\end{figure*}

		Quantum computing shows immense promise for enabling simulations of complex many-body quantum systems for materials science and quantum chemistry \cite{Bauer2020}.
		Solving realistic problems will require hundreds or thousands of nearly perfect quantum bits (qubits) \cite{vonBurg2021}, necessitating scalable implementations.
		Superconducting circuits are one leading implementation for qubits that fulfill this criterion \cite{Krantz2019}.
		Due to finite qubit coherence times and control accuracy, quantum error correction, based, for example, on the surface code \cite{Fowler2012, Versluis2017}, will be needed, requiring millions of physical qubits (depending on qubit error rates and noise model assumptions) \cite{Gidney2021, Lee2021, vonBurg2021}.
		Fabricating this quantity of qubits remains a formidable engineering challenge and will require innovative techniques such as flip-chip bonding to combine multiple planar (single-layer) chips \cite{Rosenberg2017, OBrien2017a, Foxen2018} and superconducting through-substrate vias to suppress package modes \cite{Vahidpour2017,Mallek2021}.
		While air-bridge \cite{Steffen2013,Chen2014} crossings can overcome some routing challenges \cite{Mukai2020}, planar devices will remain limited to a maximum routing density set by the acceptable crosstalk between closely spaced signal traces \cite{Besedin2018}.
		Instead, the multi-chip approach will ultimately prove more fruitful since circuit elements (qubits, couplers, readout resonators, \emph{etc}.)\ can be placed on separate chips which have optimized fabrication procedures or even different material platforms \cite{Brecht2016, Rosenberg2020}.

		In flip-chip bonding, two patterned devices are joined face-to-face by bumps of superconducting metal, typically indium due to its ductility and facile cold-welding \cite{Rosenberg2017, OBrien2017a, Foxen2018}.
		The inter-chip spacing, $d$, is a key parameter since it affects the frequencies of resonant features, the impedance matching between different signal lines, and the capacitive and inductive coupling rates between elements (such as for qubit--qubit couplers or the qubit--readout resonator coupling) \cite{Gold2021}.
		Values of $d$ between \SI{5}{\um} and \SI{10}{\um} are typical, with smaller separations increasing the inter-chip capacitance (\emph{cf}.\ a parallel-plate capacitor) and hence the coupling rates at the expense of increased electric field redistributions (compared to planar designs) that change device parameters like the phase velocity of transmission lines \cite{Simons2001}.

		Relative chip tilt is problematic for 3D-integrated devices since it leads to local changes in the chip-to-chip separation, $d$, and hence to local frequency shifts of device components.
		An investigation by Foxen \emph{et al.}\ found \SI{500}{\micro \radian} of tilt for an indium-based flip-chip bonding process, which corresponds to \SI{6}{\um} of separation difference across a \SI{12}{\mm} chip, a large fraction of the chip separation \cite{Foxen2018}.
		This leads to predicted local frequency shifts of several hundred \si{\MHz} (several \si{\percent}) for coplanar-waveguide (CPW) resonators when using typical dimensions (discussed in Appendix~\ref{sec:appendix:frequencyvschipsep}).
		The anticipated errors will be even larger for resonator coupling rates to qubits \cite{Gold2021} or the feedlines used for readout multiplexing since the rates depend on higher powers of the coupling capacitance.

		To avoid the tilt or deviation of the chip separation compared to the target value, Niedzielski \emph{et al.}\ and Li \emph{et al.}\ have demonstrated hard-stop spacers which mechanically support the chip \cite{Niedzielski2019, Li2021j}.
		Silicon spacers \cite{Niedzielski2019} are ideal from a process-compatibility perspective, but uniformly etching large silicon wafers without increasing surface roughness or loss rates is a significant fabrication challenge.
		Alternatively, large indium pads \cite{Li2021j} can act as spacers by significantly increasing the indium surface area and diluting the bonding force.
		Such indium pads are simple to define during the usual indium bump deposition process but their height can be difficult to control due to the substantial thickness being deposited.
		Recently, Somoroff \emph{et al.}\ have detailed a hybrid approach, where, instead of using separate indium bumps and hard spacers, they have used bumps composed primarily of aluminum with a thin coating of indium \cite{Somoroff2023}.
		This process saves the space required for dedicated spacers but still suffers from the difficulty of evaporating thick films with precise thicknesses.
		Therefore, we chose to develop a spacer process based on SU-8, which has previously been used in situations where galvanic connections are not required \cite{Satzinger2019, Connor2021}.
		Favorable properties of SU-8 spacers include: a simple fabrication process, suitability for wafer-scale processing, compatibility with standard fabrication procedures for low-loss devices, excellent height uniformity, and independent control over the chip separation.
		
		Here, we present this SU-8 spacer process for indium flip-chip bonding.
		In Section~\ref{sec:main:device-architecture}, we specify our device architecture and fabrication details.
		Then, we analyze the impact of the SU-8 spacers on inter-chip spacing and tilt in Section~\ref{sec:main:su8-performance}.
		Next, we discuss the frequency reproducibility of resonators on devices with spacers in Section~\ref{sec:main:resonator-frequencies}, before analyzing the quality factors of the resonators as a function of their geometric parameters in Section~\ref{sec:main:qfactors} and concluding in Section~\ref{sec:main:discussion}.

	\section{Device Architecture and Fabrication}
		\label{sec:main:device-architecture}

		Our multi-chip module [Fig.~\ref{fig:main:1}(a,b)] comprises a resonator chip (top) bonded to a wiring chip (bottom) with \SI{10}{\um} of nominal spacing.
		Here, the wiring chip [Fig.~\ref{fig:main:1}(c)] includes the multiplexed feedlines and wire-bond connections to the break-out printed-circuit board (PCB) while all resonators are on the top resonator chip [Fig.~\ref{fig:main:1}(d)].
		Superconducting indium bumps (\SI{25}{\um} diameter, \SI{10}{\um} thickness) mechanically support the resonator chip and galvanically join the ground planes of the two chips to suppress spurious modes.
		\SI{600}{\um} by \SI{600}{\um} by \SI{10}{\um} pads of SU-8 photoresist in the corners of the overlap area on the bottom wiring chip support the resonator chip during bump bonding and act as a mechanical stop to ensure uniform chip separation [Fig.~\ref{fig:main:1}(e)].
		Larger versions of the optical micrographs are provided in Appendix~\ref{sec:appendix:fabrication}.

		We fabricate these devices on \SI{100}{\mm} high-resistivity ($>\,$\SI{20}{\kilo \ohm \cm}) silicon wafers onto which we sputter \SI{125}{\nm} of niobium before patterning the film with \ce{SF6}-based reactive ion etching.
		Afterwards, we pattern \SI{10}{\um} of SU-8 3010 photoresist on the wiring chips to act as spacers.
		Next, we pattern a negative photoresist, thermally evaporate \SI{10}{\um} of indium on both the wiring and resonator chips, and then remove the unwanted indium by dissolving the photoresist under it in a solvent bath, lifting it off.
		After dicing, we flip the resonator chip, align it with the wiring chip using a split-prism microscope inserted between the two chips, and then bond them by compressing the indium bumps against each other at room temperature.
		To perform microwave measurements, we glue the device to a sample package and wire bond it to a PCB.
		For additional details about the fabrication process, see Appendix~\ref{sec:appendix:fabrication}.

	\section{SU-8 Spacer Performance}
		\label{sec:main:su8-performance}

		\begin{figure*}
			\centering
			\includegraphics{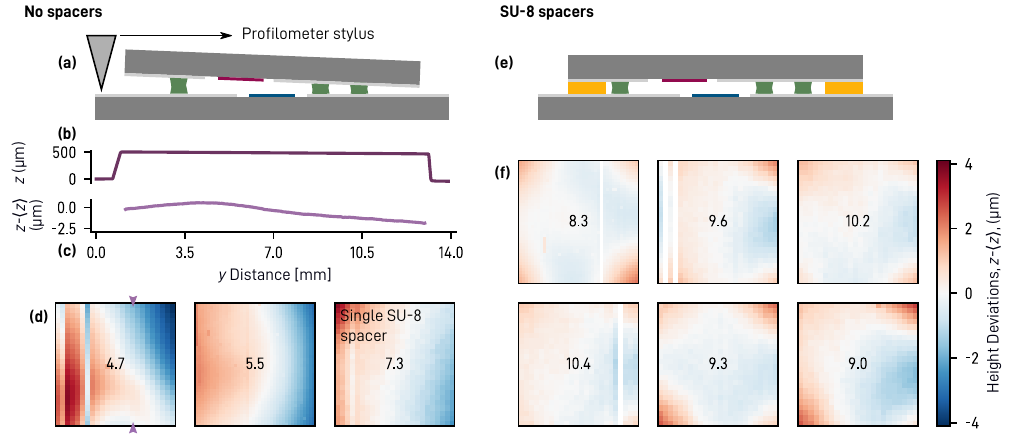}
			\caption{
				Mechanical profilometry measurements.
				(a) Schematic of a flip-chip bonded device without spacers indicating the scan direction of the mechanical profilometer stylus.
				(b) Raw profilometer data and (c) data after leveling, cropping to the top-chip region, and subtracting the mean height of the entire top-chip area.
				(d) Mechanical profilometer height maps of the modules bonded without spacers.
				The purple arrows in the leftmost panel indicate the trace presented in (c). The rightmost panel has a single spacer in the upper left corner, as indicated.
				(e) Schematic of a flip-chip bonded device with spacers and corresponding height maps (f).
				The height maps are approximately \SI{11}{\mm} by \SI{11}{\mm} in size.
				The colors represent the deviation from the mean top-chip height on each module while the number in the center of each sub-panel is the extracted average inter-chip spacing (measured value minus estimated top-chip substrate thickness).
			}
			\label{fig:main:2}
		\end{figure*}

		To comprehensively inspect the inter-chip separation, we use mechanical profilometry, where, as depicted in Fig.~\ref{fig:main:2}(a), a stylus is drawn linearly across the sample while recording the deflection, resulting in a height \emph{vs.}\ position line scan [Fig.~\ref{fig:main:2}(b)].
		We observe a step of just greater than \SI{500}{\um} from the upper surface of the bottom chip to the upper surface of the top chip, corresponding to the substrate thickness of the top chip plus the chip separation.
		We level the data based on the bottom chip and window out the top chip region, resulting in a trace [Fig.~\ref{fig:main:2}(c)] with a smoothly varying profile, with some tilt and bow (curvature), and with total deviations from the mean of around \SI{3}{\um}.

		Performing a series of such line scans, and processing the data as discussed above, we prepare the height maps presented in Fig.~\ref{fig:main:2}(d,f).
		Some line scans in the height maps are offset by several \si{\um} from adjacent ones due to measurement artifacts in the profilometer; we remove (mask) these traces, resulting in the white vertical lines in Fig.~\ref{fig:main:2}(f).
		Note that, for plotting, we subtract the mean height of the entire top-chip region for the data displayed in Fig.~\ref{fig:main:2}(c,d,f).
		Furthermore, we point out that this technique cannot distinguish chip separation from thickness variations of the top chip substrate.
		Independently, we measured the standard deviation of our wafer thicknesses at \SI{1.0}{\um} or below (discussed further in Appendix~\ref{sec:appendix:mechanicalmeasurements}).

		In devices without spacers [Fig.~\ref{fig:main:2}(d)], we observe large tilts as evidenced by the color gradient as well as deviations relative to the mean of $\pm\SI{4}{\um}$.
		Once spacers are added [Fig.~\ref{fig:main:2}(f)], tilts are substantially reduced and there are no longer large chip-separation gradients from one side of the sample to the other.
		Instead, now that large tilts are avoided, we observe bowing, with the corners raised by roughly \SI{1}{\um} and the center depressed by slightly less than that.

		For quantitative analysis (and the text values in Fig.~\ref{fig:main:2}(d,f)), we subtract the estimated top-chip substrate thickness (from independent measurements; see Appendix~\ref{sec:appendix:mechanicalmeasurements}).
		We observe a mean separation of \SI{5.8(19)}{\um} over four devices without spacers and \SI{9.6(8)}{\um} for nine devices with spacers.
		We compute the tilt for these chips by fitting a plane to the data using a least-squares method and find a mean tilt of \SI{269(151)}{\micro \radian} for the spacerless devices and \SI{76(35)}{\micro \radian} for the devices with spacers.
		Thus, based on our analysis, the spacers reduce the error from the target separation by a factor of 10 and the tilt by a factor of 3.5.
		
		Furthermore, to enable comparisons to published results \cite{Satzinger2019,Li2021j,Kosen2022}, we have also measured the chip-to-chip separation at the corners of the top chip with scanning electron microscopy (SEM).
		While chip separation information in the middle of the sample is not available without destructive techniques, we find quantitative agreement with the corners of the profilometer height maps.
		More details about the SEM measurements and the results are presented in Appendix~\ref{sec:appendix:mechanicalmeasurements}.

		Here we note that SU-8 spacers require some special care since they absorb common solvents used for resist stripping and swell up, necessitating special drying procedures and reducing the height uniformity compared to the heights immediately after spinning, developing, and baking (discussed in Appendix~\ref{sec:appendix:fabrication}).
		Comparing our measured data to literature values, the relative deviations from the target height are similar to those reported for silicon \cite{Niedzielski2019} and indium \cite{Li2021j} spacers, although the silicon spacers reported yet smaller tilts (calculated from spacer heights prior to bonding rather than measured on bonded devices).
		Additionally, while the separation and tilt errors of current spacerless processes \cite{Kosen2022} have improved compared to early reports \cite{Foxen2018}, they are still larger than for processes with spacers.
		Thus, despite minor fabrication issues, SU-8 performs comparably in practice to indium and silicon spacers.

		The bowing apparent in Fig.~\ref{fig:main:2}(f) is a concern since it could replace tilt as the dominant source of local frequency errors.
		The observed bowing could be the result of the geometry of the flip-chip bonder, elastic compression of the SU-8 spacers which leads to inelastic compression of the indium \footnote{The Young's modulus of SU-8 is nearly two orders of magnitude smaller than that of silicon.}, and the current layout of spacers located only at the edges of the resonator chip.
		The flip-chip bonder and spacer placement can easily be adjusted, but compression of the SU-8 requires adapting to lower-force indium bonding or replacement by a less-compressible spacer material.

	\section{Resonator Frequency Targeting}
		\label{sec:main:resonator-frequencies}

		Having shown that the SU-8 spacers improve our chip separation and planarity targeting, we next verify that this results in reproducible parameters for microwave circuits.
		In particular, we investigate resonator frequencies since they are important for fast, multiplexed readout circuits \cite{Heinsoo2018} in which readout resonators must be matched to Purcell filters within tens of \si{\MHz} ($\approx \SI{0.5}{\percent}$ relative accuracy).

		While standard planar CPWs have electrical properties ideally determined entirely \footnote{Assuming that the CPW dimensions are significantly smaller than the substrate thickness and distance to any metallic enclosure.} by the permittivity of the substrate and the ratio, $w/(w + 2s)$, between the center conductor width, $w$, and the gap width, $s$, the electrical properties of 3D-integrated CPWs depend additionally on the layout of conducting and dielectric features on, and the distance, $d$, to the opposite chip \cite{Simons2001}.
		Typical planar CPWs have center conductor widths, $w \approx \SI{10}{\um}$ \cite{Goeppl2008}, greater than or equal to the attainable chip separations, $d$, with evaporated indium bumps.
		Since the impedance changes are particularly acute when $w \gtrapprox d$, we utilize a smaller $w = \SI{5}{\um}$ and adapt the gaps on either side, $s$, to target a \SI{50}{\ohm} impedance.
		This balances reduced precision of lithographic processes, increased kinetic inductance \cite{Watanabe1994}, and increased losses due to the greater electric field strength \cite{Gao2008} against separation-dependent properties and compactness.

		In addition, we can cover the chip opposite the CPW with varying amounts of metal, which will change the boundary conditions for the electric field and hence influence the microwave properties (phase velocity and losses); see Appendix~\ref{sec:appendix:frequencyvschipsep} for concrete examples and discussion of the limiting cases.
		To our knowledge, the behavior as a function of material facing the CPW has not been investigated to date in this context.
		Here, we compare resonators on the top chip facing a solid metal film on the bottom chip in the region opposite the CPW (\emph{metal facing}) and resonators where the metal has been etched away during device fabrication in a \SI{140}{\um} wide strip centered across from the CPW to expose the dielectric beneath (\emph{dielectric facing}).
		For dielectric-facing devices, we leave small strips of metal ($\approx \SI{10}{\um}$ wide) in this etched region to connect the ground planes and avoid spurious modes; see Fig.~\ref{fig:main:1}(c) and the device design renders in Fig.~\ref{fig:appendix:samplerenders} for more information.

		\begin{figure}
			\centering
			\includegraphics{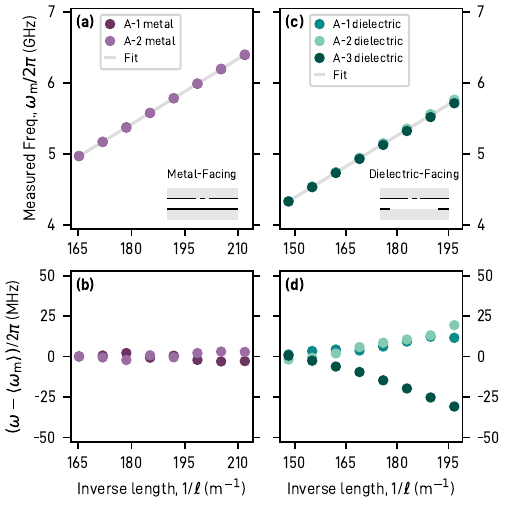}
			\caption{
				3D-integrated resonator frequency reproducibility.
				(a) Measured frequencies of a set of metal-facing resonators \emph{vs.}\ inverse physical length from multiple copies of the same design (plotted separately as A-$n$, see legend) with \SI{5}{\um} wide center conductors.
				The lower-right inset is a schematic cross-section of the layout of the CPW line with silicon in gray and niobium in black (not to scale).
				(b) Deviations between the measured frequencies and the mean measured per-resonator frequency for the data in (a).
				(c) Measured frequency and (d) deviations from mean per-resonator frequency \emph{vs}.\ inverse length from the dielectric-facing resonators.
				The light gray lines in (a,c) are best-fit lines to a simple model discussed in the text.}
			\label{fig:main:3}
		\end{figure}

		We designed samples with two feedlines of eight weakly coupled quarter-wavelength CPW resonators with frequencies staggered in \SI{200}{\MHz} increments from \SI{4.5}{\GHz} to \SI{6.5}{\GHz} and coupling quality factors of approximately \num{2e6}.
		One feedline features metal-facing CPWs while the other has dielectric-facing CPWs, see Fig.~\ref{fig:main:1}(c).
		Since the feedline is located on the wiring chip while the resonators are on the other chip, we couple them with inter-chip parallel-plate capacitors.
		Further details about the sample designs are available in Appendix~\ref{sec:appendix:sample-details}.

		We cooled the resonators down to approximately \SI{15}{\milli \kelvin} in a dilution refrigerator and measured the complex scattering parameters of the resonators with a vector network analyzer (VNA).
		Additional information about the measurement setup is available in Appendix~\ref{sec:appendix:microwave-setup}.
		We extracted the resonator frequency at a drive power which provides good signal-to-noise ratio and where the resonator does not show nonlinear behavior using a fitting technique which is robust to impedance mismatches \cite{Probst2015}.

		In a first measurement, we verify that the fundamental resonance frequency scales with the physical length of the resonator using a sample (design A) with \SI{5}{\um}-wide CPW center conductors.
		For the simplest case (the metal-facing resonators), we plot the measured resonance frequencies against the inverse physical length, $1/\ell$, of the resonator in Fig.~\ref{fig:main:3}(a) and observe a linear scaling.
		This indicates that these CPWs behave as expected, with a resonant frequency given by $v_\mathrm{ph}/4\ell$ where $v_\mathrm{ph}$ is the effective phase velocity of this particular geometry and dielectric.
		We fit the mean per-resonator measured frequency to a simple analytical model that accounts for the additional frequency shift due to the coupling to the feedline (presented in Appendix~\ref{sec:appendix:analyticalresonatormodel}) by a least-squares method and extract a phase velocity of $v_{\mathrm{ph,m}}^{\mathrm{fit}} = \SI{1.182e8}{\m \per \s}$ for this geometry.
		This $v_\mathrm{ph}$ may be process specific, since the phase velocity of a CPW will depend on details such as the metal film thickness, the degree of over-etching of the substrate, or oxide films on the various surfaces.

		Furthermore, we analyze the reproducibility of the measured frequencies by comparing the nominally identical resonators on two or three copies of this design.
		We plot the difference between the measured frequencies and the mean of all measured frequencies of each resonator on the sample in Fig.~\ref{fig:main:3}(b) and observe small deviations with a mean absolute value of \SI{3}{\MHz} or $\approx \SI{0.05}{\percent}$ between the same resonators on different copies of the device (16 resonators on two devices).
		This indicates excellent reproducibility for the metal-facing CPW geometry.

		Next, we investigate the case of CPWs with a \SI{5}{\um}-wide center conductor and dielectric facing the CPW line.
		We again find frequencies proportional to $1/\ell$ [Fig.~\ref{fig:main:3}(c)] and fit a phase velocity of $v_{\mathrm{ph,d}}^{\mathrm{fit}} = \SI{1.175e8}{\m \per \s}$.
		For these resonators, we observe a mean absolute frequency difference between different copies of the same resonator of \SI{15}{\MHz} [Fig.~\ref{fig:main:3}(d); 24 resonators on three devices].
		This deviation is dominated by copy A-3 which had slightly damaged spacers and a rightward tilt with a magnitude twice as large as that of the other two copies (the height maps of samples A-1 to A-3 are plotted in order from left to right in the lower row of Fig.~\ref{fig:main:2}(f)).
		Since the resonators decrease in length from left to right along each feedline, the rightward tilt of the top chip on copy A-3 tends to shift the frequency of the shorter resonators downward, as in Fig.~\ref{fig:main:3}(d).
		Excluding this device, we calculate a mean absolute frequency difference of \SI{3}{\MHz} across 16 resonators on two devices.
		
		Combining the results of both metal- and dielectric-facing resonators, we find a mean absolute deviation of the frequency deviations of \SI{12}{\MHz} and a maximum deviation between two nominally identical resonators of \SI{50}{\MHz} (\SI{3}{\MHz} mean and \SI{7}{\MHz} maximum difference when excluding sample A-3).
		Although these data are from a single fabrication round, this frequency spread is at or below the typical wafer-to-wafer frequency variation we observe for planar devices fabricated using similar methods but without spacers or indium bumps \footnote{For 28 resonators on eight devices, we find a mean standard deviation of resonator frequencies of \SI{23}{\MHz}.}, indicating that with intact SU-8 spacers, the flip-chip bonding process is not expected to limit the frequency repeatability of weakly coupled CPW resonators.
		Resonators with larger coupling capacitors may show greater sensitivity to interchip spacing deviations since the spurious capacitances to ground that shift the frequency (see Appendix~\ref{sec:appendix:analyticalresonatormodel}) will depend on $d$.

		Given good frequency reproducibility, it is useful to model the phase velocity of such resonators to target specific frequencies in future designs.
		Typical approaches include analytical techniques such as conformal mapping \cite{Wen1969,Gevorgian1995,Simons2001} or finite-element (FEM) simulations \cite{Kosen2022}.
		We find that conformal mapping calculates phase velocities of $v_{\mathrm{ph,m}}^{\mathrm{CM}} = \SI{1.215e8}{\m \per \s}$ ($v_{\mathrm{ph,d}}^{\mathrm{CM}} = \SI{1.185e8}{\m \per \s}$) for the metal-facing (dielectric-facing) CPWs
		while 3D radio-frequency (RF) FEM simulations produce phase velocities of $v_{\mathrm{ph,m}}^{\mathrm{FEM}} = \SI{1.233e8}{\m \per \s}$ ($v_{\mathrm{ph,d}}^{\mathrm{FEM}} = \SI{1.205e8}{\m \per \s}$).
		These values are within \SI{5}{\percent} of the measured values in the worst case, and a large portion of the remaining error is likely attributable to kinetic inductance \cite{Frunzio2005, Watanabe1994}.
		In general, accurately modeling the absolute frequencies of CPW resonators is challenging due to: (i) geometrical effects resulting from fabrication such as the metal thickness and the over-etch into the substrate; (ii) kinetic inductance of the Cooper pairs in the superconductor which depends on the film thickness and penetration depth; and (iii) parasitic effects missing in the equivalent model.
		Complicating matters further, (i) and (ii) may vary spatially across the wafer.
		
	\section{Resonator Quality Factors}

		\label{sec:main:qfactors}

		Having analyzed the reproducibility of resonator frequencies and modeled them, we next evaluate the impact of the additional process steps introduced for the 3D-integrated samples on the low-photon number internal quality factors of the CPW resonators.
		Quality factors at this power are important since this is the regime in which qubits are operated \cite{McRae2020a}.
		For the $w = \SI{5}{\um}$ CPWs, we find a mean internal quality factor at single photon levels of \num{0.5(1)e6}, with no statistically significant difference between metal-facing and dielectric-facing resonators [Fig.~\ref{fig:main:4}(a)].
		We discuss the calculation of the resonator internal photon number in Appendix~\ref{sec:appendix:resonator-internal-photon-number}.

		\begin{figure}
			\centering
			\includegraphics{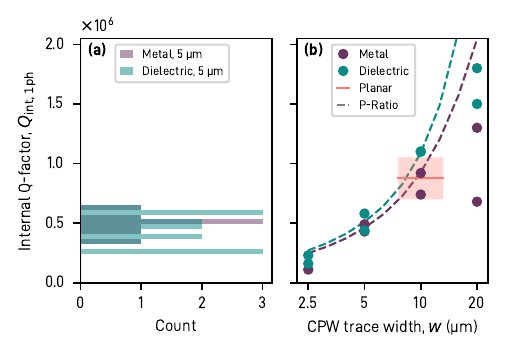}
			\caption{
				Analysis of 3D-integrated resonator quality factors.
				(a) Distribution of single-photon internal quality factors for 24, $w=\SI{5.0}{\um}$, CPW resonators across two devices.
				(b) Internal quality factor \emph{vs}.\ CPW center conductor width, $w$, for both metal and dielectric-facing resonators.
				A logarithmic $x$-axis scale is used for visual clarity.
				The pink region indicates the mean and one standard deviation interval for the planar CPW resonators with $w=\SI{10}{\um}$ and \emph{aq.}\ HF treatment prior to cooldown.
				The dashed lines are the result of participation-ratio analysis of the different geometries.
			}
			\label{fig:main:4}
		\end{figure}
		
		Considering a second sample (design B) where the CPW center conductor width is set to \SIlist{2.5; 5; 10; 20}{\um} for different resonators while adapting the gap size to approximately maintain a \SI{50}{\ohm} impedance according to the conformal mapping model (resulting in proportionally larger gaps at large $w$), we observe that the internal quality factors are strongly affected by CPW size, scaling by a factor of $\approx \num{5}$ from $w=\SI{2.5}{\um}$ to $w=\SI{20}{\um}$ [Fig.~\ref{fig:main:4}(b)].
		We expect larger $w$ and gaps to reduce the field strength and thus the participation of lossy interfaces, resulting in increased quality factors.
		Indeed, the observed dependence agrees qualitatively with numerical participation-ratio analysis of the geometries (discussed in Appendix~\ref{sec:appendix:participation-ratio-analysis}).
		Importantly, the $w = \SI{10}{\um}$ resonators have a single-photon internal quality factor of $\approx \num{1e6}$, which is comparable to airbridge- and qubit-free, hydrogen-fluoride (HF) treated planar resonators with identical $w$ produced in our lab using similar fabrication techniques but without SU-8 spacers or indium bumps.
		Thus, the \num{5e5} quality factors reached by the $w=\SI{5}{\um}$ resonators on both the A and B samples are likely limited by our decision to use a smaller $w$ compared to planar designs to reduce the sensitivity of transmission line properties on chip separation rather than additional steps in the 3D integration process.

	\section{Discussion}
		\label{sec:main:discussion}

		To improve parameter reproducibility of indium flip-chip bonded superconducting microwave circuits, we have developed an SU-8 spacer process.
		We showed that the spacers reduce the mean chip separation error as well as tilt by factors of 10 (to a mean of \SI{0.4(8)}{\um}) and 3.5 (to a mean of \SI{76(35)}{\micro \radian}) respectively.
		Furthermore, we demonstrated the ability of the profilometry technique to characterize the entire bonded area by uncovering bowing which is not visible with SEM measurements of the chip corners or pre-bonding measurements of the spacer heights.
		Additional investigation of such bowing is required, particularly the impact of additional spacers.
		SU-8 spacers have advantages due to the simplicity and accessibility of their fabrication process.
		Based on our measurements, SU-8 spacers perform comparably to silicon and indium spacers.
		However, further study will be needed to evaluate their qubit compatibility and the best approach, particularly given their propensity for absorbing solvents during standard cleaning steps used in state-of-the-art qubit fabrication.

		We also verified that the reproducible chip separation results in CPW resonator frequencies with statistical device-to-device frequency deviations at the typical wafer-to-wafer variations, which should enable multiplexed readout circuits including Purcell filters in future work.
		For 3D-integrated CPWs with a \SI{5}{\um}-wide center conductor, we found that standard techniques do not model the phase velocities with sufficient precision for \emph{ab initio} device design and will need to be refined in future studies, particularly by investigating kinetic inductance in these geometries.
		To directly test the impact of spacer height variations on resonator frequency shifts, we could vary the height of the spacers, which would also help to improve our models.

		Measurements of the resonator quality factors show that the flip-chip bonding process preserves the low-loss material interfaces of a similar planar fabrication process at the level of quality factors of \num{5e5}.
		Resonators with wider center conductors reached higher internal quality factors in exchange for potentially increased sensitivity to chip separation deviations.
		We did not find significant differences between metal- and dielectric-facing resonators, indicating that we can use either based on design convenience.
		In particular, due to their reduced phase velocity sensitivity on chip spacing deviations, dielectric-facing resonators may be preferred at the cost of physically longer resonators due to their lower phase velocity.
		More work is needed to investigate alternative geometries (microstrip-like) to avoid the quality-factor penalties of narrow CPW lines and preserve separation-independent properties.
	
	\section{Acknowledgments}

		We thank Stephan Paredes for his generous assistance with the indium evaporations.
		We also thank the staff of the ETH Zürich FIRST and the Binnig and Rohrer Nanotechnology Center cleanrooms for their assistance in maintaining the facilities used to prepare the devices discussed in this work.
		The finite-element simulations were performed using the ETH Zürich Euler cluster.

		The authors acknowledge financial support by the EU Flagship on Quantum Technology H2020-FETFLAG2018-03 project 820363 OpenSuperQ, by the Office of the Director of National Intelligence (ODNI), Intelligence Advanced Research Projects Activity (IARPA), via the U.S.\ Army Research Office grant W911NF-16-1-0071, by the National Center of Competence in Research Quantum Science and Technology (NCCR QSIT), a research instrument of the Swiss National Science Foundation (SNSF), by the SNFS R'equip grant 206021-170731, by the Swiss State Secretariat for Education, Research and Innovation (SERI) under contract number UeM019-11, and by ETH Zürich.
		The views and conclusions contained herein are those of the authors and should not be interpreted as necessarily representing the official policies or endorsements, either expressed or implied, of the ODNI, IARPA, or the U.S.\ Government.

	\section{Author Contributions}

		G.J.N., J.-C.B., and C.E.\ planned the experiments.
		G.J.N.\ designed and G.J.N.\ and M.K.\ fabricated the devices.
		G.J.N.\ and D.P.\ performed the mechanical measurements.
		G.J.N.\ performed the microwave measurements and analyzed all data.
		L.M.\ performed the participation-ratio analysis.
		G.J.N.\ wrote the manuscript with input from all authors.
		C.E.\ and A.W.\ supervised the work.

	\appendix

	\section{Frequency Dependence on Chip Separation}
		\label{sec:appendix:frequencyvschipsep}
	
\begin{figure}[t]
			\centering
			\includegraphics{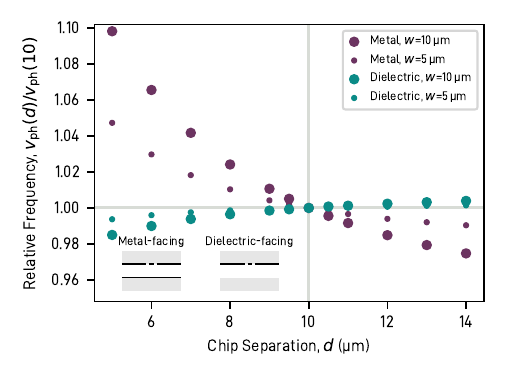}
			\caption{
				Simulated relative frequency shifts as a function of inter-chip spacing for dielectric- and metal-facing CPW resonators.
			}
			\label{fig:appendix:frequencyvsheight}
		\end{figure}

		Since it is challenging to accurately produce a sequence of chips over a large range of target separations, we turn to simulations to numerically explore the influence of chip separation on the frequency of CPW resonators.
		We consider a center conductor width, $w$ of \SI{10}{\um} and gaps of \SI{5.5}{\um} to ground on either side on top of \SI{525}{\um} of silicon with a relative permittivity, $\epsilon = 11.45$ \cite{Krupka2006}, and assume that there is either a \SI{525}{\um} thick piece of silicon (dielectric-facing) or a sheet of metal (metal-facing) a distance $d$ away on the opposite chip.
		We calculate the phase velocity, $v_{\mathrm{ph}}(d)$ using 3D RF FEM simulations and plot the relative frequency shift of a resonator as the ratio $v_{\mathrm{ph}}(d)/v_{\mathrm{ph}}(\SI{10}{\um})$ in Fig.~\ref{fig:appendix:frequencyvsheight}.

		The dielectric-facing CPW resonator shifts downwards in frequency with decreasing chip separation due to the increasing electric field participation in the dielectric (increasing the effective permittivity $\epsilon_{\mathrm{eff}}$ and lowering $v_{\mathrm{ph}}$).
		The metal-facing CPW resonator shifts upwards in frequency with decreasing chip separation due to the increasing electric field participation in the vacuum region between the center trace and the metal film above (increasing $v_{\mathrm{ph}}$).
		Decreased chip separation relative to the target value results in larger deviations than increased separation, and metal-facing CPWs experience far larger frequency shifts with changing $d$ than dielectric-facing ones due to the different boundary conditions.

		In both cases, we see that $<\SI{1}{\um}$ deviations around the target distance of \SI{10}{\um} result in frequency deviations below \SI{1}{\percent}, but large negative deviations ($\approx \SI{5}{\um}$) result in frequency shifts of at least a few percent or hundreds of \si{\MHz} for resonant frequencies around \SI{5}{\GHz}.
	
		Furthermore, we have simulated CPWs with $w=\SI{5}{\um}$ and gaps of \SI{3.24}{\um} (\SI{3.14}{\um}) for the metal-facing (dielectric-facing) resonators, which show reduced frequency shifts with changing $d$ compared to the $w=\SI{10}{\um}$ CPWs, motivating our choice of $w=\SI{5}{\um}$ for the devices in this work.

		The absolute phase velocities for the $w=\SI{5}{\um}$ CPWs at $d=\SI{10}{\um}$ are presented in the main text.
		For the $w=\SI{10}{\um}$ CPWs at $d=\SI{10}{\um}$, we calculate $v_{\mathrm{ph,m}}^{\mathrm{FEM}} = \SI{1.275e8}{\m \per \s}$ ($v_{\mathrm{ph,d}}^{\mathrm{FEM}} = \SI{1.190e8}{\m \per \s}$) for the metal-facing (dielectric-facing) CPWs.

	\section{Fabrication Details}
		\label{sec:appendix:fabrication}

		\subsection{Niobium Deposition and Patterning}

			We fabricate the devices discussed in this work on \SI{100}{\mm}, (100)-orientation, high-resistivity ($>\,$\SI{20}{\kilo \ohm \cm}) intrinsic float-zone silicon wafers (\emph{Topsil GlobalWafers A/S}).
			We clean the wafers for \SI{5}{\minute} in a $1:1$ mixture of \SI{25}{\percent} ammonium hydroxide (\ce{NH4OH}) and \SI{30}{\percent} hydrogen peroxide (\ce{H2O2}) at \SI{60}{\degreeCelsius} to remove organic contaminants before stripping the native silicon oxide for \SI{60}{\s} in a \SI{7}{\percent} solution of hydrofluoric acid (\ce{HF}) in water at room temperature and then rinsing with de-ionized (DI) water.
			Within \SI{20}{\minute}, we place the cleaned wafers into the load-lock of an ultra-high-vacuum (base pressure $<\SI{1e-7}{\pascal}$) magnetron sputtering system (\emph{AJA International Inc.}), where we sputter $\approx \SI{125}{\nm}$ of niobium from a \SI{100}{\mm} niobium target (\SI{99.99}{\percent}, \emph{ACI Alloys, Inc.}) in a face-to-face geometry over \SI{300}{\s} with a \SI{25}{\sccm} flow of \ce{Ar} resulting in a chamber pressure of $\approx \SI{1}{\pascal}$.
			Before venting to the atmosphere, we expose the fresh niobium film to nitrogen gas for \SI{15}{\minute}.

			After unloading the wafer, we clean it with sonication in a \SI{50}{\degreeCelsius} bath of isopropanol to remove particles before spinning AZ 5214E (EU) photoresist (\emph{Microchemicals GmbH}) (\SI{45}{\s} of spinning at \SI{4000}{\per \minute} followed by a \SI{60}{\s} bake at \SI{105}{\degreeCelsius} on a hotplate).
			We expose the wafers in contact mode with a mask aligner (\emph{EV Group}, EVG 620NT) at an exposure dose of \SI{60}{\milli \joule \per \cm \squared} of an even mix of \SIlist{365;405;420}{\nm} light-emitting-diode (LED) illumination.
			After exposure, we develop the resist for \SI{60}{\s} in AZ 726 MIF (\emph{Microchemicals GmbH}) followed by \SI{60}{\s} of rinsing in DI water and a spin rinse and dry.
			We etch the now-exposed niobium film in a reactive-ion etcher (\emph{Oxford Instruments}, Plasmalab 80 Plus) using \ce{SF6} chemistry, a chamber pressure of \SI{9e2}{\Pa}, a flow rate of \SI{5}{\sccm}, and an RF power of \SI{100}{\W} for approximately \SI{240}{\s} using the reflectivity of the surface to a helium-neon laser to determine the end of the etch.
			We strip the resist for at least \SI{120}{\minute} in \SI{80}{\degreeCelsius} \emph{N}-Methyl-2-pyrrolidone (NMP) followed by \SI{10}{\minute} NMP, acetone, and then isopropanol sonication at \SI{50}{\degreeCelsius}.

			After stripping, we measure step heights from the niobium surface to the silicon below of approximately \SI{155}{\nm} in the center of the wafer increasing to \SI{185}{\nm} at the edge of the wafer using mechanical profilometry.
			Using our measured niobium thickness of approximately \SI{125}{\nm} in the center of the wafer, this indicates over-etches into the silicon of approximately \SI{30}{\nm} in the center and \SI{60}{\nm} at the edge of the wafer.

		\subsection{SU-8 Patterning}

			Before starting the patterning of the SU-8, we clean the wafers for \SI{60}{\s} in a \SI{7}{\percent} HF solution at room temperature and then rinse with DI water.
			We spin SU-8 3010 (\emph{Kayaku Advanced Materials, Inc.}) for \SI{60}{\s} at \SI{3000}{\per \minute} and then allow the film to rest for \SI{5}{\minute} on the spinner before soft baking for \SI{180}{\s} at \SI{95}{\degreeCelsius}.
			We expose the SU-8 on a mask aligner (EVG 620NT) in contact mode at an exposure dose of \SI{200}{\milli \joule \per \cm \squared} of \SI{365}{\nm} LED illumination and perform post-exposure bakes at \SI{65}{\degreeCelsius} for \SI{60}{\s} and then \SI{300}{\s} at \SI{95}{\degreeCelsius} on a vacuum hotplate.
			We develop for \SI{90}{\s} in mr-Dev 600 and then wash several times in alternating isopropanol and mr-Dev 600 baths until no residues remain.
			To improve the mechanical resilience of the SU-8 spacers, we next hard bake at \SI{180}{\degreeCelsius} for \SI{900}{\s}.
			Finally, we perform mechanical profilometer (\emph{Bruker Corp.}, DektakXT) measurements of the niobium and SU-8 thicknesses.

			We note that, since the baking temperature of typical electron-beam-lithography (EBL) resists used for Josephson junction fabrication are above the melting temperature of indium (\SI{156}{\degreeCelsius}), 
			EBL needs to be performed prior to indium deposition.

		\subsection{Indium Patterning}

			Before starting the indium patterning, we clean the wafers for \SI{60}{\s} in a \SI{7}{\percent} HF solution at room temperature (\SI{45}{\s} for wafers with SU-8) and then rinse with DI water.
			We start by spinning AZ nLOF 2070 (\emph{Microchemicals GmbH}) for \SI{1}{\s} at \SI{3000}{\per \minute} with \SI{1}{\s} ramps on either side before allowing the wafer to rest on the spinner for \SI{300}{\s} with the lid open.
			Then, we soft bake for \SI{30}{\s} at \SI{100}{\degreeCelsius} before removing the edge bead with a few \si{\mL} of propylene glycol methyl ether acetate (PGMEA) while spinning at \SI{500}{\per \minute} and then \SI{30}{\s} at \SI{1500}{\per \minute} once the PGMEA has been applied.
			We bake for \SI{360}{\s} at \SI{100}{\degreeCelsius} on a vacuum hotplate and then expose on a mask aligner (EVG 620NT) in contact mode at an exposure dose of \SI{110}{\milli \joule \per \cm \squared} of an even mix of \SIlist{365;405;420}{\nm} LED illumination.
			Next, we perform a post exposure bake at \SI{110}{\degreeCelsius} for \SI{60}{\s} on a vacuum hotplate before developing in AZ 826 MIF (\emph{Microchemicals GmbH}) for \SI{90}{\s} and then rinsing in DI water.
			We then load the wafer into a thermal evaporator (\emph{Plassys Bestek}, MEB550S) and perform an \emph{in-situ} argon ion mill for \SI{300}{\s} at a beam voltage of \SI{500}{\V}, a beam current of \SI{35}{\mA}, and an argon flow of \SI{6}{\sccm}.
			Without breaking vacuum, we evaporate $\approx \SI{10}{\um}$ of indium at \SI{10}{\nm \per \second} with the wafer temperature held at approximately \SI{4}{\degreeCelsius} in a water-cooled chuck.
			To complete the indium deposition, we lift-off the indium on top of the resist film in \SI{50}{\degreeCelsius} acetone over \SI{120}{\minute}.

		\subsection{Dicing, Cleaning, and SU-8 Drying}

			With the indium patterning completed, the individual dies are finished.
			To prepare for dicing, we spin AZ 4533 (\emph{Microchemicals GmbH}) resist for \SI{60}{\s} at \SI{1000}{\per \minute} before baking for \SI{90}{\s} at \SI{80}{\degreeCelsius} to protect the front surface of the wafer.
			After dicing, we clean individual chips with isopropanol and acetone to remove the resist followed by \SI{15}{\minute} in \SI{80}{\degreeCelsius} NMP to remove resist residues, then \SI{60}{\minute} in \SI{50}{\degreeCelsius} acetone and bake for at least \SI{12}{\hour} in a vacuum oven at \SI{50}{\degreeCelsius} and \SI{2e4}{\Pa} to remove the solvents from the SU-8 spacers.

			Immediately after spinning, exposure, development, and baking, the mean SU-8 3010 spacer height is \SI{10.00(4)}{\um}.
			However, after immersing in warm solvents (in particular, NMP), the SU-8 spacers increase in thickness up to \SI{25}{\percent}, which we counteract with a solvent-exchange and drying procedure (detailed above).
			Immediately prior to bonding, the SU-8 spacer height is approximately \SI{10.2(2)}{\um}.

		\subsection{Flip-chip Bonding and Packaging}

			Then we flip-chip bond the bottom and top chips together in a flip-chip bonder (\emph{Smart Equipment Technology Corp.\ SA}, FC150).
			The bonder uses an autocollimator to ensure that the bottom and top chips are parallel prior to bonding and a split-prism microscope inserted between the chips to align them laterally.
			We calibrate the parallelism of the arm and chuck, align the autocollimator, and finally align the microscope so that it points to the same locations on the chuck and arm.
			After inserting the chips, we align the bottom and top chips in five axes (lateral position and rotation as well as the two rotation axes for parallelism) and press them together at room temperature with a force between \SIrange{10}{40}{\N \per \mm \squared} of indium (\SI{20}{\N \per \mm \squared} typical).
			Unlike Ref.~\cite{Foxen2018}, we do not use atmospheric plasma cleaning to remove the indium oxides prior to bonding.

			After flip-chip bonding, we glue (GE 7031 Varnish) the module onto an oxygen-free high thermal conductivity (OFHC) copper base with a microwave printed circuit board (PCB) attached.
			Next, we connect launchers on the PCB and bottom chip using a manual wedge-type wire bonder (\emph{West Bond, Inc.}, 7476E) with \SI{25}{\um} diameter aluminum wire.
			Finally, we close the sample package with a 6082 aluminum alloy lid and vacuum bag the sample for transport from the cleanroom to the laboratory where it is installed in a cryostat.

		\subsection{Additional imagery}

			\begin{figure*}
				\centering
				\includegraphics{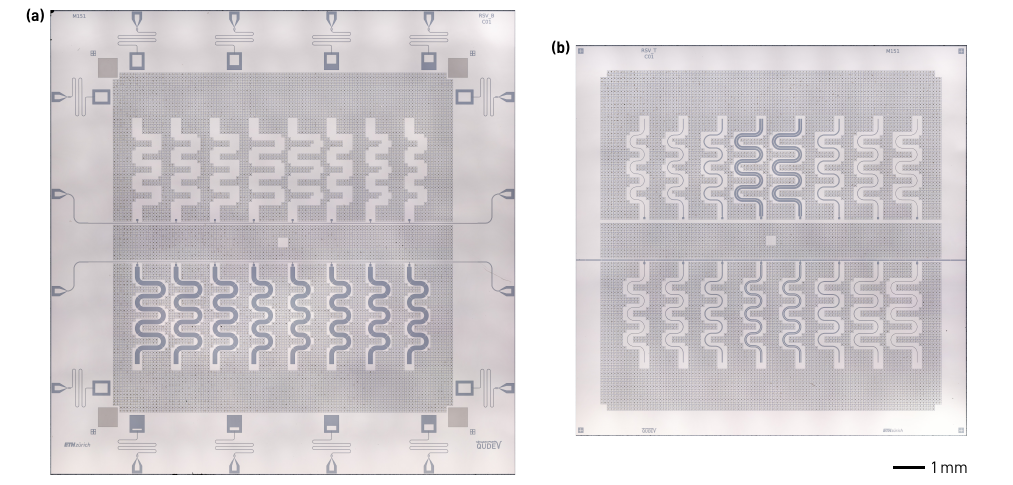}
				\caption{
					High-resolution composite micrographs of a copy of sample design B with varied CPW center conductor width.
					(a) Bottom chip.
					(b) Top chip.
				}
				\label{fig:appendix:samplemicrographs}
			\end{figure*}

			We present additional composite micrographs of a copy of the design B with varied CPW center conductor width in Fig.~\ref{fig:appendix:samplemicrographs}.
			These composite images have been created by aligning and merging numerous images taken with a microscope including corrections for lighting non-uniformity (Hugin) before performing curve adjustments to increase contrast (GNU image manipulation program).
			The micrographs in the main text [Fig.~\ref{fig:main:1}(c,d,e)] are desaturated versions of these micrographs which have been artificially colored.

	\section{Mechanical Measurements}
		\label{sec:appendix:mechanicalmeasurements}

		\subsection{Mechanical Profilometry}

			\begin{figure}[b]
				\centering
				\includegraphics{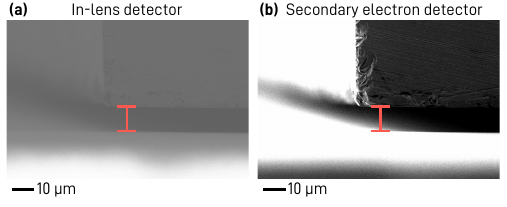}
				\caption{
					Edge-on SEM micrographs of one corner of a flip-chip bonded device:
					(a) In lens detector, backscattered electrons.
					The cursor height, corresponding to the measured separation and indicated by the pink bar, is \SI{10.90}{\um}.
					(b) Secondary electron detector.
					The cursor height is \SI{10.80}{\um}.
				}
				\label{fig:appendix:semheightanalysis}
			\end{figure}

			We measure each bonded module in the mechanical profilometer (\emph{Bruker Corp.}, DektakXT) starting from the lower-right corner of the bottom chip.
			The dektak scans first vertically from bottom to top and then repeats such scans from the right edge of the bonded device until the left edge.
			We level the data by fitting a plane to bottom chip region in the complete dataset with a least-squares method and then subtracting this plane from the entire dataset.
			Next, we mask out individual vertical line scans that are clearly offset from the rest of the measurements to avoid skewing the results.

			Finally, we subtract the substrate thickness from the measured top-chip heights.
			We first estimate the thickness of the substrates by accurately measuring the thickness of other wafers from the same batches using a precision micrometer (\emph{Mitutoyo Corp.}, MDH-25MB).
			We used two different types of wafers for these devices: double-side polished for the mechanical test samples, and single-side polished for the resonator samples.
			We find that our double-side polished wafers have a mean thickness of \SI{505.9(10)}{\um} and that our single-side polished wafers have a mean thickness of \SI{525.2(4)}{\um}.
			We thus select the data for the top-chip region and subtract the appropriate substrate thickness to arrive at the extracted chip separations.

		\subsection{Scanning-Electron Microscopy}

			We image the corners of the top chip of a bonded device edge-on with two different detectors in the scanning-electron microscope (SEM) and measure the gaps manually using changes in contrast to detect the bottom-chip and top-chip edges (see Fig.~\ref{fig:appendix:semheightanalysis})
			Since the bottom-chip edge is visible only by differences in local contrast due to depth-of-focus, there is some ambiguity in these measurements which we attempt to reduce by using two different detectors.
			Additionally, since we are only with a few degrees of perpendicular to the edge, we estimate that these measurements have an uncertainty of a few hundred \si{\nm}.
			We then average the chip separations from both detectors into a single value for that corner.
			The average separation for the module is the mean of the corner separations.
			To compute the tilt, we utilize the methodology of Ref.~\cite{Kosen2022}, \emph{i.e.}\ we compute the inverse tangent of the chip separation difference divided by the lateral distance for all six corner pairs on the device and quote the largest value.
			A tilt extracted from fitting a plane to the measured data is typically lower than these worst-case local tilts.

			Without spacers, we calculate a mean per-module corner separation of \SI{6.1(2)}{\um} and mean per-module worst-case tilt of \SI{450(200)}{\micro \radian} across four bonded modules.
			This calculation is similar to that of Ref.~\cite{Kosen2022} and we find comparable, although slightly worse, values.
			With spacers, we extract a separation of \SI{11.0(3)}{\um} and a tilt of \SI{62(26)}{\micro \radian} over nine modules, a factor of 4 reduction in chip separation deviations and a factor of 7 reduction of planarity errors.
			All devices in Fig.~\ref{fig:main:3}(f) excluding the one in the lower right corner are included in this set of nine devices.
			We find that the corner separations extracted from SEM measurements are consistent with the mechanical profilometry.
			The larger inter-chip separation results from the SEM method are likely caused by the observed bowing.

	\section{Sample Details}
	\label{sec:appendix:sample-details}

		\begin{figure*}
			\centering
			\includegraphics{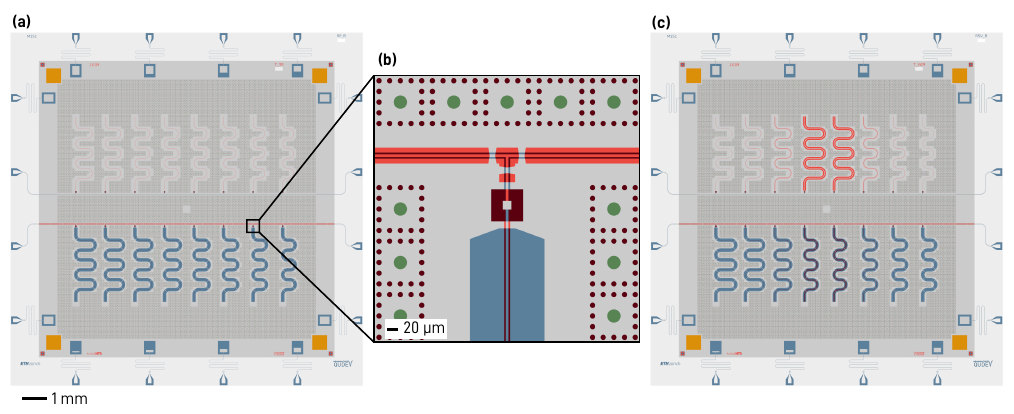}
			\caption{
				Renders of the device designs:
				(a) Sample A with uniform $w$ and $s$ for each CPW resonator type.
				(b) Detail of an inter-chip coupler between a feedline and a resonator.
				(c) Sample B with swept $w$ and $s$ for each CPW resonator type.
				In all images, niobium is shown as light gray on the bottom chip and darker gray on the top chip.
				Areas of exposed silicon on the bottom chip are rendered in light blue while equivalent areas on the top chip are rendered in magenta.
				Indium bumps are rendered in green and SU-8 spacers are rendered in yellow.
			}
			\label{fig:appendix:samplerenders}
		\end{figure*}
		
		All results in this work are based on \SI{14.3}{\mm} by \SI{14.3}{\mm} bottom chips and \SI{12.0}{\mm} by \SI{12.0}{\mm} top chips.
		We present renders of the device designs in Fig.~\ref{fig:appendix:samplerenders}.
		Devices with SU-8 spacers have four \SI{600}{\um} by \SI{600}{\um} rectangles of SU-8 placed on the bottom chip just within the outline of the top chip.
		They are sized such that, assuming a Young's modulus of \SI{2}{\giga \pascal} \cite{KayakuSU83000}, we expect a compression of only \SI{7}{\percent} when using a bonding force of \SI{200}{\N}.

		The indium bumps are \SI{10}{\um} high, \SI{25}{\um} diameter, and have a pitch of \SI{100}{\um}.
		The samples have a total number of bumps ranging from \num{8297} to \num{11096}, resulting in a total area of between \SI{4}{\square \mm} and \SI{5.4}{\square \mm} of indium.

		\begin{table}
			\centering
			\caption{
				CPW resonator parameters of design A.
				Resonator indices increase from left to right along the feedline; \numrange{0}{7} refer to the metal-facing resonators on the upper feedline while \numrange{8}{15} refer to the dielectric-facing resonators on the lower feedline.
				$w$: center conductor width, $s$: gap width, $\ell$: physical length, $\langle \omega_{\mathrm{m}} \rangle$: mean measured resonance frequency.
			}
			\label{tab:appendix:sample:a:rescpwparams}
			\begin{tabular}{S S S S S} \toprule
				{Index} & {$w$ [\si{\um}]} & {$s$ [\si{\um}]} & {$\ell$ [\si{\um}]} & {$\langle \omega_{\mathrm{m}} \rangle/2\pi$ [\si{\GHz}]}  \\ \midrule
				 0 &  5.00 &  3.24 & \num{6049.8} & \num{4.9729} \\
				 1 &  5.00 &  3.24 & \num{5816.1} & \num{5.1724} \\
				 2 &  5.00 &  3.24 & \num{5599.7} & \num{5.3748} \\
				 3 &  5.00 &  3.24 & \num{5398.7} & \num{5.5797} \\
				 4 &  5.00 &  3.24 & \num{5211.6} & \num{5.7865} \\
				 5 &  5.00 &  3.24 & \num{5037.0} & \num{5.9921} \\
				 6 &  5.00 &  3.24 & \num{4873.7} & \num{6.1990} \\
				 7 &  5.00 &  3.24 & \num{4720.5} & \num{6.3988} \\ \midrule
				 8 &  5.00 &  3.14 & \num{6749.1} & \num{4.3347} \\
				 9 &  5.00 &  3.14 & \num{6447.9} & \num{4.5386} \\
				10 &  5.00 &  3.14 & \num{6172.4} & \num{4.7407} \\
				11 &  5.00 &  3.14 & \num{5919.4} & \num{4.9423} \\
				12 &  5.00 &  3.14 & \num{5686.2} & \num{5.1448} \\
				13 &  5.00 &  3.14 & \num{5470.6} & \num{5.3458} \\
				14 &  5.00 &  3.14 & \num{5270.7} & \num{5.5467} \\
				15 &  5.00 &  3.14 & \num{5082.8} & \num{5.7475} \\
				\bottomrule
			\end{tabular}
		\end{table}

		The dimensions of the CPW resonators for sample designs A are listed in Table~\ref{tab:appendix:sample:a:rescpwparams}.
		The resonators are coupled to the feedline with parallel-plate capacitors made of overlapping \SI{16}{\um} by \SI{16}{\um} square pads with a uniform \SI{22}{\um} gap to ground on all sides [see Fig.~\ref{fig:appendix:samplerenders}(b)].
		In electrostatic simulations (\emph{Ansys, Inc.}, Maxwell 2022 R1), we compute a capacitance of approximately \SI{0.44}{\fF} between the pads, and surplus capacitances to ground (compared to a coplanar waveguide of the equivalent length) of approximately \SI{0.6}{\fF} on the resonator and feedline side.
		We define the physical length of the resonators, $\ell$, to start from the center of the square coupling capacitor pad.

	\section{Microwave Measurement Setup}
		\label{sec:appendix:microwave-setup}

		\begin{figure}
			\centering
			\includegraphics{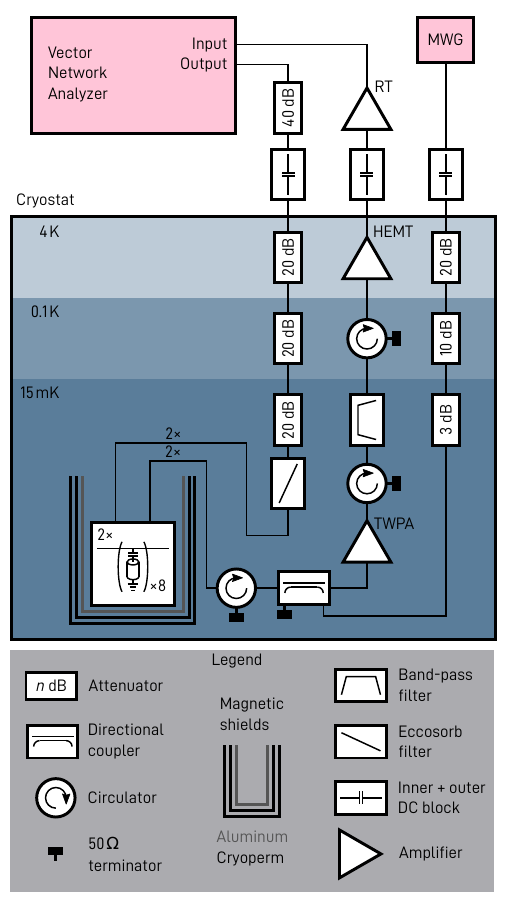}
			\caption{
				Diagram of our measurement setup for the cryogenic microwave measurements.
			}
			\label{fig:appendix:wiringdiagram}
		\end{figure}

		The signal path (see Fig.~\ref{fig:appendix:wiringdiagram}) begins at a vector network analyzer (\emph{Agilent Technologies}, N5230C) before passing through a \SI{40}{\dB} attenuator and then an inner and outer DC block.
		The cryostat (\emph{Bluefors Oy}, LD250) features further \SI{20}{\dB} attenuators at the \SI{4}{\kelvin}, \SI{100}{\mK}, and \SI{15}{\mK} stages of the input line before a final custom coaxial Eccosorb filter (\emph{Laird plc.}, Eccosorb CR-110).
		The reasoning behind this choice of attenuators is presented in Ref.~\cite{Krinner2019}.
		The sample is enclosed in a package with a copper base and aluminum lid and then placed inside a high-purity aluminum magnetic shield surrounded by two high permeability nickel-alloy shields (\emph{Magnetic Shields Ltd.}, Cryophy).
		The output line features an isolator (\emph{Low Noise Factory}, ISIS4\_12A), a \SI{20}{\dB} directional coupler, a traveling wave parametric amplifier (TWPA) (\emph{MIT Lincoln Labs}), another LNF isolator, and then a band-pass filter on or below the base temperature stage.
		The output line then has an additional circulator at the \SI{100}{\mK} stage and a high-electron mobility transistor (HEMT) amplifier (\emph{LNF}, LNC4\_8A) at the \SI{4}{\K} stage.
		Outside the cryostat, the output line has an inner and outer DC block and the room-temperature amplification chain consisting of: an ultra-low-noise amplifier (ULNA), a low-pass filter, a \SI{10}{\dB} attenuator, a low-noise-amplifier (LNA) , a \SI{3}{\dB} attenuator, and another inner and outer DC block.

		Our measurements were performed without pumping the TWPA (\emph{i.e.}\ with it off) to avoid saturation effects at high probe powers and frequency-dependent gain that might distort the resonator lineshapes.

	\section{Analytical Resonator Model}
		\label{sec:appendix:analyticalresonatormodel}

		\begin{figure}
			\centering
			\includegraphics{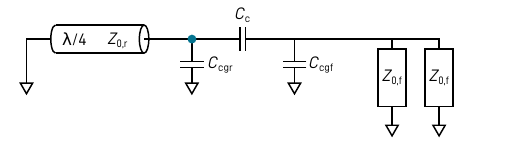}
			\caption{
				Circuit representation of the coupling circuit between the $\lambda$/4 resonators and the feedline.
			}
			\label{fig:appendix:resonatorcouplingcircuit}
		\end{figure}

		Here, we analyze an electrical model of a CPW resonator capacitively coupled to a feedline to determine the resonant frequency under this additional loading and to find a model to extract the CPW phase velocity from resonator frequency measurements as a function of resonator length.
		As shown in Fig.~\ref{fig:appendix:resonatorcouplingcircuit}, we consider our resonator as a $\lambda/4$ transmission line of impedance $Z_{0,\mathrm{r}}$ and bare resonance frequency $\omega_0$ connected to a two-ended feedline of impedance $Z_{0,\mathrm{f}}$ by a coupling capacitance $C_{\mathrm{c}}$.
		We include a parasitic capacitance to ground $C_{\mathrm{cgr}}$ ($C_{\mathrm{cgf}}$) on the resonator (feedline) side and expect that the spurious capacitance to ground on the resonator side will lower the resonant frequency.

		We write out the total impedance of the circuit about the selected node (the blue dot in Fig.~\ref{fig:appendix:resonatorcouplingcircuit}) and extract the loaded resonant frequency from the poles of this impedance.
		Setting $C_{\mathrm{cgf}}=0$ to simplify notation (it only contributes at high order) and assuming $(C_\mathrm{c} Z_{0,\mathrm{f}} \omega/2)^2$ is small, we find the resonance condition:
		\begin{equation}
			\tan{\left(\frac{\pi}{2} - \frac{\ell}{v_\mathrm{ph}} \frac{\omega_\mathrm{r}}{2\pi}\right)} - \left( C_\mathrm{c} + C_\mathrm{cgr}\right) Z_{0, \mathrm{r}} \omega_\mathrm{r} = 0
		\end{equation}
		where we have substituted $\omega_0 = v_\mathrm{ph}/4\ell$ for a $\lambda/4$ resonator with phase velocity, $v_\mathrm{ph}$, and physical length $\ell$.
		Assuming that the capacitive frequency shift is small, so $\omega_\mathrm{r} \approx \omega_0$, we can expand the tangent and arrive at a solution for $\omega_\mathrm{r}$:
		\begin{equation}
			\omega_\mathrm{r} = \frac{\pi}{2} \frac{1}{ \frac{\ell}{v_\mathrm{ph}} + \left( C_\mathrm{c} + C_\mathrm{cgr} \right) Z_{0, \mathrm{r}} }.
		\end{equation}
		We fit this model to our measured resonator frequencies using the design lengths (Table~\ref{tab:appendix:sample:a:rescpwparams}) with the phase velocity and $b = \left( C_\mathrm{c} + C_\mathrm{cgr} \right) Z_{0, \mathrm{r}}$ as free parameters.

	\section{Resonator Internal Photon Number}
		\label{sec:appendix:resonator-internal-photon-number}

			We compute the internal photon number of the resonator as a function of applied power using the following formula, which can be easily derived \cite{Bruno2015}
			\begin{equation}
				n_{\mathrm{int}} = 2 \frac{\kappa P_{\mathrm{app}}}{\hbar \omega_0 (\kappa + \gamma)^2}
			\end{equation}
			where $\kappa$ is the  coupling rate to the feedline, $P_{\mathrm{app}}$ is the power applied at the input port of the sample, $\hbar$ is the reduced Planck constant, $\omega_0$ is the resonant frequency, and $\gamma$ is the internal loss rate.
			We subtract the input line attenuation measured at room temperature from the power supplied by the VNA to arrive at the power applied to the sample input which results in uncertainty of a few \si{\dB}.

	\section{Participation-Ratio Analysis}
		\label{sec:appendix:participation-ratio-analysis}

		To understand the influence of the CPW geometry on losses, we numerically analyze the electric-field distribution, focusing on the fraction of the electric field energy (the \emph{participation ratio}) stored in thin layers on the surfaces of the CPW and substrate representing amorphous surface oxides which are believed to host two-level systems (TLS) that induce loss \cite{McRae2020a}.
		This technique is widely used to correlate device geometry with losses and thus quality factors \cite{Wenner2011, Wang2015}.

\begin{figure}
			\centering
			\includegraphics{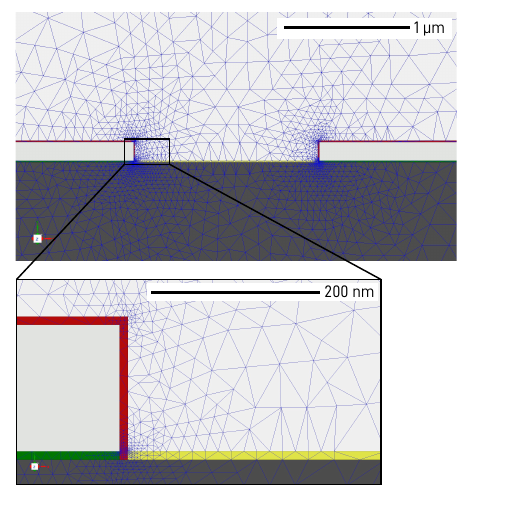}
			\caption{
				Meshing of the participation-ratio simulations.
				Gray is bulk substrate, green is metal---substrate interface, red is metal---vacuum interface, and yellow is substrate---vacuum interface region.
				The mesh triangles are outlined in blue.
				The inset presents a detailed view of the \SI{10}{\nm} interface region near the lower corner of the CPW metal where the electric-field is highest.
			}
			\label{fig:appendix:participationratiomeshing}
		\end{figure}

		We partition a two-dimensional, side-cut slice of the chosen CPW geometry into metal, substrate, and vacuum bulk regions as well as metal---substrate (MS), metal---vacuum (MV), and substrate---vacuum (SV) interface regions \cite{Wenner2011, Calusine2018}.
		We follow standard practice and treat the interface regions as having a thickness of \SI{10}{\nm} and a dielectric constant of $\epsilon = \num{10}$ \cite{Wenner2011, Calusine2018, Woods2019}.
		We calculate the electric field distribution using an electrostatic solver (\emph{Ansys, Inc.}, Maxwell 2022 R1) configured to perform adaptive meshing steps until the change in participation ratios of the regions outlined above is below \SI{1}{\percent} from one iteration to the next.
		We repeat such simulations for all CPW geometries of device B and interpolated values of $w$ in between the measured geometries.
		See Fig.~\ref{fig:appendix:participationratiomeshing} for a diagram of the different regions considered as well as the final mesh for a $w=\SI{2.5}{\um}$, $s=\SI{1.49}{\um}$ CPW line.

		Since the participation ratios in the different lossy interfaces are highly correlated (meaning that they scale together with changes in $w$ or $s$) \cite{Woods2019}, we are unable to use the calculated participation ratios along with the measured internal quality factors to extract loss tangents of the interfaces (for this, devices with extreme geometries [\emph{e.g.}\ isotropically etched trenches in the CPW gaps] would be required, as in Ref.~\cite{Woods2019}).
		Instead, we follow a simplified procedure to create the participation-ratio curves in Fig.~\ref{fig:main:4}, described here.
		For each geometry, we compute the total interface participation, $p_\Sigma$:
		\begin{equation}
			p_\Sigma(w,x) = \sum_{i}{p_i(w,x)}
		\end{equation}
		where $p_i$ is the participation ratio in one of the three lossy interfaces (MS, MV, or SV), $w$ is the CPW center conductor width, and $x$ is either metal- (m) or dielectric-facing (d).
		Then, we compute relative Q-factors:
		\begin{equation}
			Q_{\mathrm{pr}}(w,x) = Q_{\mathrm{meas}}(\SI{5}{\um},x) \frac{p_\Sigma(\SI{5}{\um}, x)}{p_\Sigma(w, x)}
		\end{equation}
		using the mean single-photon internal quality factors for the $w=\SI{5}{\um}$ resonators of each type on device B.

		Compared to the measured data, this procedure overestimates the quality factors at large $w$, likely since it does not separate the power-dependent (two-level system) and power-independent losses which add in parallel, \emph{e.g.} as done in Ref.~\cite{Calusine2018}.
		Furthermore, the participation-ratio curves in Fig.~\ref{fig:main:4} overestimate the quality factors at small $w$ slightly, which could be due to our equal weighting of all interfaces.

	\section{Finite-Element Electrostatic Simulations}

		We simulate the capacitance matrix of the coupling capacitors discussed in this work using a finite-element electrostatic solver (\emph{Ansys, Inc.}, Maxwell 2022 R1).
		We model each conductor as \SI{125}{\nm}-thick perfect electrical conductor.
		We assign voltage excitations to the ground planes and capacitor pads and then simulate until the change in total energy from one iteration to the next is below \SI{0.1}{\percent} for a minimum of two converged passes.
		The simulation volume of \SI{1000}{\um} by \SI{1000}{\um} by \SI{1060}{\um} is significantly larger than the $\approx \SI{60}{\um}$ capacitor dimensions (\emph{cf}.\ Sec.~\ref{sec:appendix:sample-details}).
	
	\bibliography{references}

\end{document}